\documentclass[12pt]{article}
\usepackage{times}
\usepackage{geometry}
\usepackage[utf8]{inputenc}
\usepackage{enumitem,amssymb}
\usepackage{ragged2e}
\newlist{thematic}{itemize}{8}
\setlist[thematic]{label=$\square$}
\usepackage{pifont}
\usepackage{natbib,graphicx}

\begin{document}
\raggedright
\huge
Astro2020 Science White Paper \linebreak

A Formaldehyde Deep Field \linebreak
\normalsize

\noindent \textbf{Thematic Areas:} \hspace*{60pt} $\square$ Planetary Systems \hspace*{10pt} $\square$ Star and Planet Formation \hspace*{20pt}\linebreak
$\square$ Formation and Evolution of Compact Objects \hspace*{31pt} $\square$ Cosmology and Fundamental Physics \linebreak
  $\square$  Stars and Stellar Evolution \hspace*{1pt} $\square$ Resolved Stellar Populations and their Environments \hspace*{40pt} \linebreak
  \makebox[0pt][l]{$\square$}\raisebox{.15ex}{\hspace{0.1em}$\checkmark$}    Galaxy Evolution   \hspace*{45pt} $\square$             Multi-Messenger Astronomy and Astrophysics \hspace*{65pt} \linebreak
  
\textbf{Principal Author:}

Name:	Jeremy Darling
 \linebreak						
Institution:  University of Colorado
 \linebreak
Email:  jeremy.darling@colorado.edu
 \linebreak
Phone:  303 492 4881
 \linebreak
 

\begin{figure}[h]
\centering
\includegraphics[width=1.0\textwidth,trim=0 0 0 20]{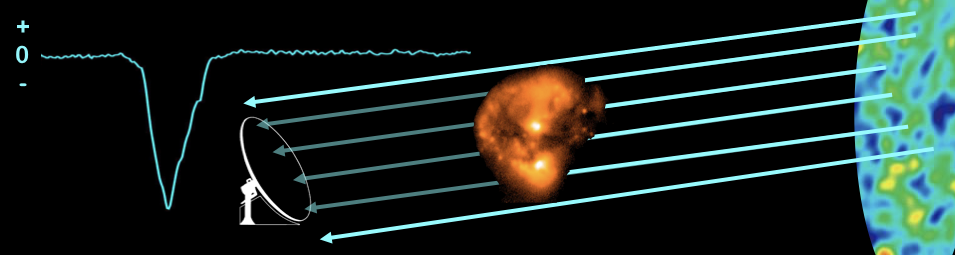}
\end{figure}

\textbf{Abstract:}
Formaldehyde (H$_2$CO) is often observed at centimeter wavelengths as an absorption line against
the cosmic microwave background (CMB).  This is possible when energy level populations are cooled
to the point where line excitation temperatures fall below the local CMB temperature.  Collisions with molecular
hydrogen ``pump'' this 
anti-maser excitation, and the cm line ratios of H$_2$CO provide a measurement of the local H$_2$ density.  H$_2$CO 
absorption of CMB light provides all of the benefits of absorption lines (no distance dimming) but none of the drawbacks:  the CMB
provides uniform illumination of all molecular gas in galaxies (no pencil beam sampling), and all galaxies lie in front of the CMB  --- no fortuitous 
alignments with background light sources are needed.   A formaldehyde deep field (FDF) would therefore provide a blind, mass-limited
survey of molecular gas across the history of star formation and galaxy evolution.  
Moreover, the combination of column density and number density 
measurements may provide geometric distances in large galaxy samples and at higher redshifts than can be done
using the Sunyaev Zel'dovich effect in galaxy clusters.  We present a fiducial FDF that would span redshifts $z=0$--7 and
provide H$_2$CO line ratios to measure $n({\rm H}_2)$ for $z > 0.45$.\footnote{Portions of this science white paper were 
adapted from \citet{darling2018} with permission from the publisher.}

\pagebreak

\setcounter{page}{1}

\vspace{-10pt}
\section{Background}

\vspace{-6pt}
The formaldehyde molecule (H$_2$CO) has peculiar non-thermal excitation properties in the physical conditions typical of 
star-forming regions.
Similar to molecules that can be induced to form masers via population inversion through a pumping mechanism, 
H$_2$CO is often ``pumped''  into an anti-inverted state by collisions with molecular hydrogen (H$_2$).  
Anti-inversion is an over-population of a lower-energy state compared to thermal, forming a kind of ultra-cold 
anti-maser that can absorb cosmic microwave background (CMB) photons if the line excitation temperature drops below
the local CMB temperature.
\citet{townes1997} glibly called this effect the ``dasar'' ---  darkness amplification by stimulated absorption of radiation\footnote{Note that darkness is not in fact amplified, nor is absorption stimulated.}.   Unlike maser excitation, which requires special local conditions, 
this anti-inversion of H$_2$CO is nearly ubiquitous in the Galaxy and in external galaxies \citep[e.g.,][]{ginsburg2011,mangum2013} 
and seems to be the natural state of the molecule for a wide range of physical conditions \citep[e.g.,][]{darling2012}.  
Moreover, H$_2$CO is observed wherever CO is present and is not strictly a high density molecular gas tracer.  The anti-inverted
$K$-doublet lines are not optically thick and their ratio is a measure of the H$_2$ number density (via collisional pumping).  
The dasar effect can thus be used to make a cosmological census of molecular gas mass and gas density, 
independent of redshift (as described below).

\hspace{10pt}   
In what follows we examine the feasibility of and science enabled by a formaldehyde deep field (FDF).  
We assume a flat cosmology with $H_0 = 70$~km~s$^{-1}$~Mpc$^{-1}$, $\Omega_m = 0.3$, and $\Omega_\Lambda = 0.7$.

\vspace{-10pt}
\subsection{Formaldehyde Anti-Inversion}\label{subsec:anti-inversion}

\begin{figure}[t]
\centering
\includegraphics[width=0.6\textwidth,trim=0 150 70 90,clip]{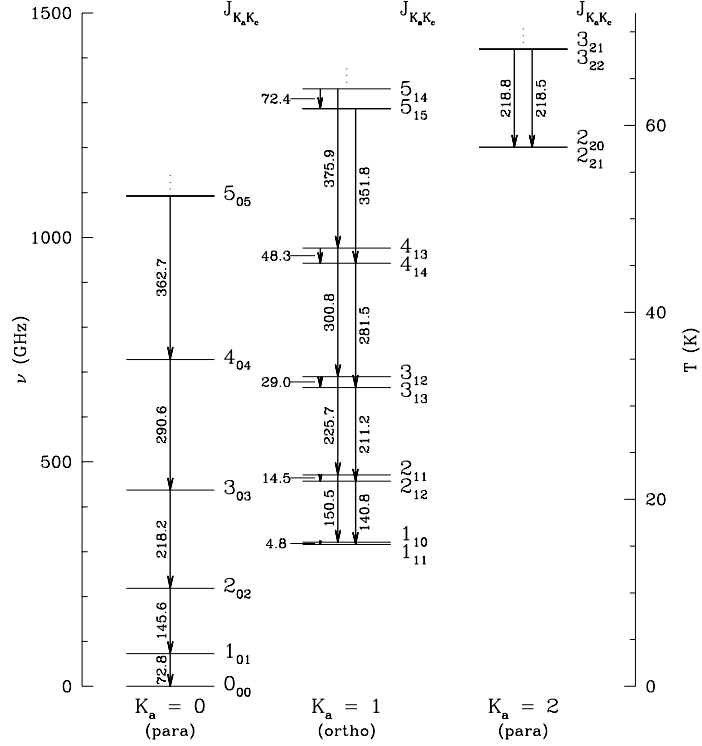}
\caption{Formaldehyde energy levels for the first three
rotation ladders and $J\leq5$ \citep[after][]{darling2012}.  Allowed transition frequencies are listed in GHz.
The anti-inverted cm lines are the $K$-doublet transitions in the $K_a=1$ ortho-H$_2$CO rotation ladder at 4.8, 14.5, 
29.0, 48.3, and 72.4 GHz (6, 2, 1, 0.6, and 0.4~cm, respectively).
\label{H2CO_levels}}
\vspace{-10pt}
\end{figure}

\vspace{-6pt}
H$_2$CO is an asymmetric top molecule with three rotation quantum numbers:  the total rotation $J$ and the rotation
about two axes, $K_a$ and $K_c$.
Each rotation state specified by $J$ and $K_a$ has two possible $K_c$ states, known as ``$K$-doublet''
splitting (the exception is the $K_a= 0$ rotation ladder of para-H$_2$CO; see Figure \ref{H2CO_levels}).
H$_2$ collisions overpopulate
the lower energy states of these $K$-doublet rotation states, creating excitation temperatures below the local cosmic microwave background (CMB) temperature by roughly 1--2~K.  The observational signature of this anti-inversion is absorption against the CMB in the centimeter wavelength $K_a = 1$ ortho-H$_2$CO lines primarily at 1, 2, and 6 cm (29.0, 14.5, and 4.8 GHz).

\hspace{10pt}     
This effect is strongest for the lowest energy cm lines and is insensitive to the local gas kinetic temperature.  
The observed line excitation temperature
is also insensitive to the local CMB temperature (which scales from its value of $T_0 = 2.73$~K today as $T_{\rm CMB}(z) =  (1+z)\,T_0$).  The anti-inversion 
favors a fairly wide range of H$_2$ density, roughly $10^2$--$10^5$~cm$^{-3}$, and the cm line ratios indicate the local gas 
number density.  The physics of H$_2$CO excitation and radiative transfer has been fairly well-studied in an extragalactic and
cosmological context:  \citet{mangum2008} and \citet{mangum2013} derived physical conditions
from observations of H$_2$CO in nearby star-forming galaxies, \citet{zeiger2010} studied H$_2$CO anti-inversion 
in the gravitational lens B0218+357 at $z=0.68$, and \citet{darling2012} performed detailed modeling of
H$_2$CO excitation versus redshift.

\vspace{-10pt}
\subsection{Absorption of CMB Light}

\vspace{-6pt}
 The absorption of CMB photons by H$_2$CO implies that the line strength in beam-matched observations is independent of distance.  
And for $z \gtrsim 1$, beam-matching is no longer a strong function of distance because angular sizes become flat for $z\simeq 1$--3
and grow thereafter (Figure \ref{fig:angsize}).  The unusual circumstances presented by H$_2$CO anti-inversion have 
some compelling consequences:  
\begin{enumerate}

\item Absorption lines do not require fortuitous alignment of the object of interest with an illuminating light source.  The CMB lies behind every
galaxy and therefore every galaxy with molecular gas may be studied in H$_2$CO absorption.  

\item Unlike traditional absorption line studies, the illuminating ``beam'' is not a pencil beam that samples a subset of the intervening 
galaxy or gas cloud.  The CMB provides an illuminating screen that is uniform to parts in $10^5$ in the CMB rest frame.  All gas is 
sampled in a manner similar to emission line observations (but absorption does not diminish with distance).  

\item The consequence of the above two points and the distance-independent nature of absorption lines is that it is possible 
to survey all H$_2$CO gas in the universe in a mass-limited fashion, provided one can beam-match to molecular gas regions in 
galaxies while achieving sub-Kelvin surface brightness sensitivity.

\item H$_2$CO absorption lines provide the column density of gas, and H$_2$CO line ratios provide the local gas number density.  This 
implies that Sunyaev-Zel'dovich-like distance measurements are possible using the molecular gas in galaxies \citep{darling2012}.  
In contrast to S-Z measurements of the X-ray gas in clusters that extend to $z\sim1$, the H$_2$CO geometric distance measurement 
may be possible in galaxies up to $z\sim6$.  While individual gas-rich galaxies are not ``spherical cows,'' inclinations can be measured and ensembles of galaxies with random orientations may be combined in redshift slices to obtain reliable distances.  
\end{enumerate}
Given these consequences of the peculiarities of the H$_2$CO molecule, one can therefore consider a formaldehyde deep field
(FDF):  a blind, gas mass-limited survey of star-forming galaxies across the history of cosmic star formation.

\vspace{-10pt}
\section{Key Observational Requirements}

\vspace{-6pt}
\subsection{Angular Resolution and Redshift Coverage}

\begin{figure}[t!]
  \includegraphics[width=1.0\textwidth,trim=5 15 10 0]{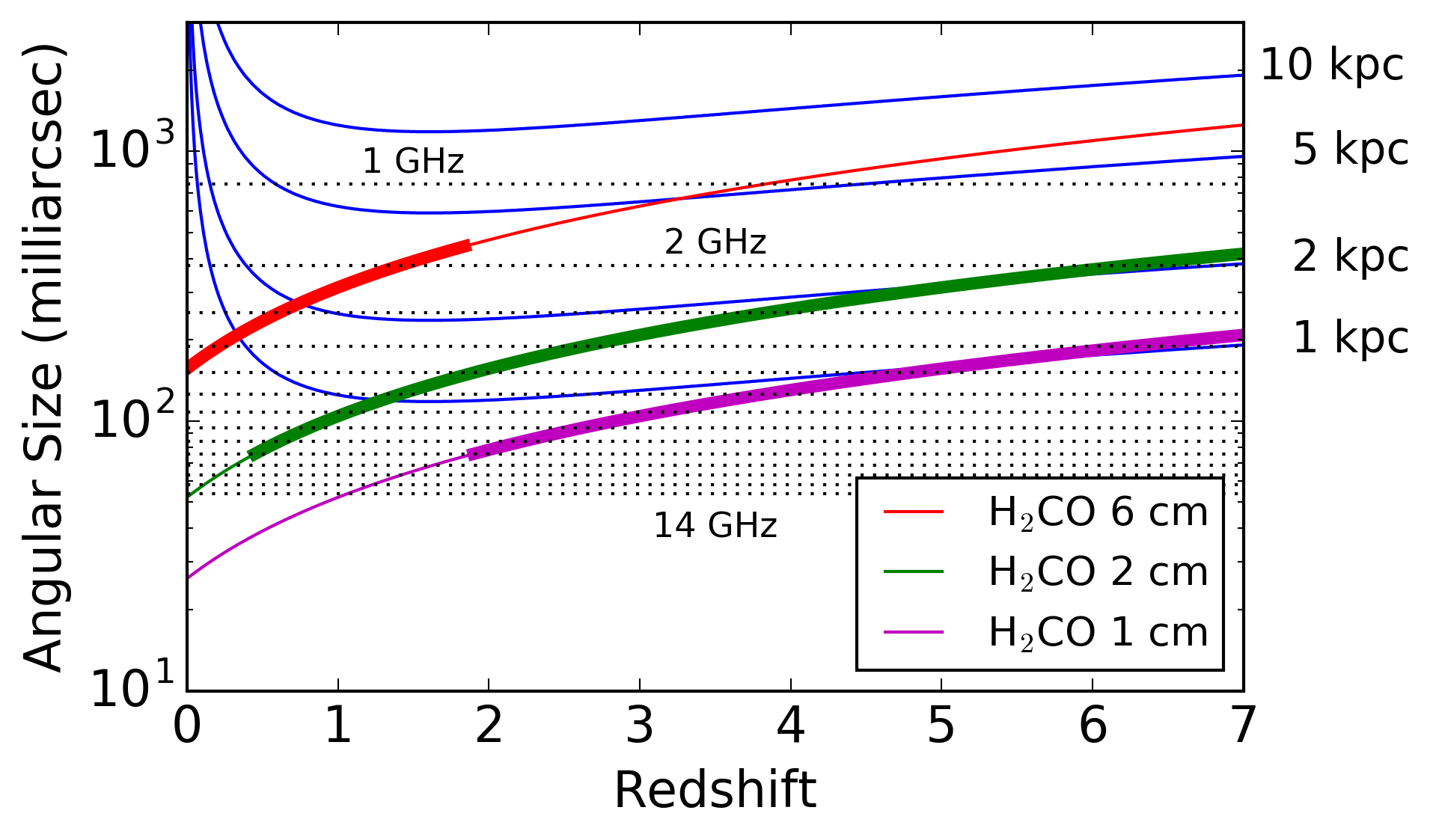}
\caption{
Angular size of various physical size scales versus redshift (blue lines).
The angular resolution of a radio array with 100 km baselines is also plotted for 
the first three CMB-absorbing H$_2$CO lines  (red, green, and magenta for 
the 6, 2, and 1 cm lines, respectively).
Black dotted lines indicate observed frequencies in 1 GHz increments, from 1 GHz
(top) to 14 GHz (bottom).  
When the angular resolution track lies
below a galaxy size scale (of a disk or starburst nucleus, 
for example), then it is possible to resolve or beam-match that scale.
For example, 5 kpc scales can always be beam-matched in all three lines
down to $\sim$1.1~GHz (where the 6 cm line crosses the 5 kpc track).  
In most cases, beam dilution will not be significant in observations of
star-forming galaxies or starburst nuclei.
The bold angular resolution tracks indicate a fiducial FDF observation spanning
1.7 to 10 GHz.  In this case, line ratios can be measured for {\it all} redshifts $z > 0.45$ (where two
line tracks overlap in redshift space).}\label{fig:angsize}
\end{figure}

\vspace{-6pt}
In order to avoid beam dilution against the CMB light screen, observations should be beam-matched to the size of star-forming regions in 
galaxies. 
Figure \ref{fig:angsize}  shows angular size tracks versus redshift for 10, 5, 2, and 1 kpc scales.  It also shows the angular
resolution of a fiducial 100~km radio array versus redshift for the 6, 2, and 1 cm H$_2$CO lines.
When the beam size for a given line and redshift is below an angular size track then that physical scale is resolved and can
be beam-matched.
Remarkably, a 100~km array can resolve or roughly beam-match 5 kpc at all redshifts in all of the cm lines.
Scales of 2 kpc or greater can be resolved in the 1 and 2 cm lines up to $z\sim6$.
A 100 km array can therefore beam-match the physical scales relevant to star-forming 
galaxies including compact starbursts and major mergers (but see Section \ref{sec:sensitivity} for a discussion 
of brightness temperature sensitivity).

\hspace{10pt}     
Figure \ref{fig:angsize} suggests that a carefully tailored frequency range will enable a formaldehyde deep field
to span the full history
of star formation (with uniform sensitivity to molecular gas mass, independent of gas kinetic temperature or
local CMB temperature) while providing line ratios to measure the {\it in situ} molecular gas density $n(\rm H_2)$.
Observations from 0.6 GHz to 15 GHz would provide complete coverage (redshifts and line ratios) in the 6 cm and 2 cm lines,
but this is problematic in implementation.  Sub-GHz observations are particularly susceptible to RFI and will not be
able to reach the required brightness temperature sensitivity using a reasonable collecting area (see below).  Only two
lines are needed at each redshift to measure $n(\rm H_2)$, so the upper and lower frequency bounds can be adjusted to
enable the 1 cm line to form a ratio with the 2 cm line when the 6 cm line is redshifted to its upper redshift bound (lowest
frequency).

\hspace{10pt}       
In Figure \ref{fig:angsize}, we illustrate a FDF spanning 1.7 to 10~GHz that enables line ratios for $z>0.45$.  
A low-bandwidth option could be 1--6~GHz, but line ratios would only be available for $z>1.4$ (and the 6 cm line may
be difficult to detect at 1 GHz in reasonable integration times).  It is also worth noting that the field of view of a
homogeneous radio array will vary inversely with frequency, so line ratios will only be available in the
highest frequency line's field of view (which is smallest at 10 GHz, corresponding to $z=1.9$ for the 1 cm line and $z=0.45$
for the 2 cm line).  

\hspace{10pt}       
A FDF spanning 1.7 GHz to 10~GHz  is reasonable from a radio receiver and backend correlation perspective.  For example,
feed horns that span an octave in frequency are now routine, and array correlators can produce more than 32,000 channels
at a time.  The proposed FDF would require roughly 64,000 channels.

\vspace{-10pt}
\subsection{Sensitivity}\label{sec:sensitivity}

\vspace{-6pt}
The observed line temperature depends on the line optical depth and on the
difference between the line excitation temperature and the background 
continuum brightness temperature, redshifted to the observer's reference frame.
For the H$_2$CO $K$-doublet lines, the continuum is the CMB
at the host galaxy's redshift (but may also include the host galaxy continuum):
\begin{equation}
 \Delta T_{\rm Obs}={T_{\rm x}(z)-T_{\rm CMB}(z)\over1+z}(1-e^{-\tau}).
\label{eqn:DTobs}
\end{equation}
\citet{darling2012} showed that the observed temperature decrement $\Delta T_{\rm Obs}$ is insensitive to redshift or
the local gas kinetic temperature and spans a large range in gas number density, 
$10^2$~cm$^{-3}\lesssim n({\rm H}_2) \lesssim 10^5$~cm$^{-3}$.  At low density, the temperature decrement 
approaches zero (the line excitation temperature equilibrates with the CMB), and at high density, the line excitation 
temperature thermalizes to the local gas temperature.    Typical temperature decrements $(T_{\rm x}(z)-T_{\rm CMB}(z))/(1+z)$
are $\sim$2~K for the 6 cm line and $\sim$1~K for the 1 cm line.  The detection of anti-inverted cm lines will therefore rely critically 
on the filling factor of molecular gas and the line optical depths that will combine to manifest as an effective 
optical depth.  For $\tau_{\rm eff}=0.1$, $\Delta T_{\rm Obs} \simeq 0.1$--0.2~K.   
 
\hspace{10pt} 
Interferometers have poor surface brightness sensitivity compared to filled apertures.  The surface brightness sensitivity
scales with frequency as $\nu^{-2}$ and with resolution as $\theta^{-2}$ (i.e., an array is least sensitive at low
frequencies and high resolution).  The best FDF will therefore be a compromise
between angular resolution, observed frequency, and surface brightness sensitivity.  This is complicated by the fact that the
lowest frequency line at 6 cm is typically the strongest anti-inverted H$_2$CO transition.

\hspace{10pt} 
In order to beam-match 5 kpc scales at all redshifts, the angular resolution would need to be roughly 600 mas.
At this resolution, the rms brightness temperature sensitivity of an array with an effective area of 50,000 m$^2$ (roughly
10 times the Very Large Array) is roughly 2.0 K  at 1.7 GHz and 0.1 K at 10 GHz.  This assumes a 100 hr integration and
100~km~s$^{-1}$ channels.  If one reduces the 
resolution to 1 arcsecond, which will resolve 10 kpc scales at all redshifts, the rms line brightness temperatures
become 0.9 K (1.7 GHz) and 50 mK (10 GHz).

\vspace{-10pt}
\section{A Formaldehyde Deep Field}\label{sec:fdf}

\vspace{-6pt}
Given the angular resolution, redshift coverage for single lines and line ratios, and sensitivity considerations above, 
the best compromise FDF could be:
\begin{itemize}
 \item  A 100-hour deep field pointing with a 50,000~m$^2$ array,
 \item  100 km s$^{-1}$ channels to adequately sample the velocity span of molecular gas in galaxies ($\sim$300~km~s$^{-1}$),
 \item  Full synthesis over 1.7--10~GHz, which will include the 6, 2, and 1 cm H$_2$CO lines 
spanning the molecular history of the universe, $z=0$--7, and
 \item  Angular resolution of 0.6 arcsec, enabling beam-matching to 5 kpc scales at all redshifts.
\end{itemize}
Line ratios will be available spanning nearly all of cosmic star formation history, for $z>0.45$.
Single-line detections at low redshift will have to be disambiguated using ancillary data such as photometric redshifts, but this
will be straightforward with a carefully-selected FDF location.

\vspace{-10pt}
\section{Conclusions}

\vspace{-6pt}
The next-generation Very Large Array is a fairly good match to the above criteria and would therefore be uniquely capable of observing a
formaldehyde deep field (FDF).  The FDF would provide a distance-independent
mass-limited census of molecular gas across the history of star formation and galaxy evolution.  H$_2$CO line ratios in the FDF 
will provide a measurement of the local H$_2$ gas density, and it may therefore be possible
to make geometric distance measurements over a large
redshift range based on the H$_2$CO $K$-doublet line depths and line ratios \citep{darling2012}.    
An FDF would complement flux-limited ``blind'' molecular emission line surveys \citep[e.g.,][]{pavesi2018}
and could break the usual degeneracy between molecular gas temperature and density encountered in line excitation studies.

\pagebreak


\end{document}